*Review paper*

# Data Driven Approaches to Cybersecurity Governance for Board Decision-Making - A Systematic Review

Anita Modi[1], Ievgeniia Kuzminykh[1], Bogdan Ghita[2]
[1] King's College London
[2] University of Plymouth
[3] Kharkiv National University of Radio Electronics, Ukraine
Correspondence: ievgeniia.kuzminykh@kcl.ac.uk

## Abstract

Cybersecurity governance influences the quality of strategic decision-making to ensure cyber risks are managed effectively. Board of Directors are the decisions-makers held accountable for managing this risk; however, they lack adequate and efficient information necessary for making such decisions. In addition to the myriad of challenges they face, they are often insufficiently versed in the technology or cybersecurity terminology or not provided with the correct tools to support them to make sound decisions to govern cybersecurity effectively. A different approach is needed to ensure BoDs are clear on the approach the business is taking to build a cyber resilient organization. This systematic literature review investigates the existing risk measurement instruments, cybersecurity metrics, and associated models for supporting BoDs. We identified seven conceptual themes through literature analysis that form the basis of this study's main contribution. The findings showed that, although sophisticated cybersecurity tools exist and are developing, there is limited information for Board of Directors to support them in terms of metrics and models to govern cybersecurity in a language they understand. The review also provides some recommendations on theories and models that can be further investigated to provide support to Board of Directors.

**Index words:** cybersecurity, cybersecurity governance, metrics, board of directors

## Introduction

Within the structure of a company, the Board of Directors (BoDs) oversee the investment and strategy of the organisation and, together with the executive management, bear responsibility for the successes and failures of the organisation. As part of their duties, BoDs are implicitly accountable for ensuring the organisation has a sufficiently resilient and robust cybersecurity strategy and associated infrastructure, through the information provided to them by the management and technical teams. While the reporting activities are well defined and follow the appropriate communication protocols, the decision process is subject to the BoDs understanding the technical implications; without the technical background to support them, BoDs remain unprepared for this role and not in a position to make sound decisions regarding overseeing implementation and execution of the cybersecurity strategy, issues regarding investments, budgets, organizational roles, and responsibilities.

The cybersecurity awareness of BoDs has improved in recent years through concerted efforts, from national security government organisations (NCSC, 2023) to commercial (Pescatore and Spitzner, 2017) resources. Nevertheless, research indicates that there is an identifiable gap in "understanding the cybersecurity information received" (Marchewka, 2022). Both BoDs and cybersecurity technical audiences see cybersecurity risk through different lenses and their understanding influences their perception and management of risk. Too often the lexicon used by technical audiences when communicating to BoDs is loaded with technical jargon, which subsequently results in poor decision-making (Marchewka, 2022). As a result, BoDs remain apprehensive when it comes to governing the posture of their organization's security, in spite of the core need to make well-informed decisions for cybersecurity governance (von Solms ,2016).

BoDs require adequate and efficient cyber risk and cybersecurity related information from management using intelligible, relatable, and meaningful language to make decisions relating to budgets, investments and manging risk. The introduction of metrics derived from cybersecurity or cyber risk information which are not tailored for the BoDs requirements is likely to lead to a suboptimal decision and potentially even hinder the decision process itself.

The first step in addressing the challenges faced by the BoDs is investigating the landscape of information that they are provided with, more specifically the risk measurement instruments and their results. In this paper, we aim to provide a comprehensive review of existing cybersecurity metrics, the gaps that need to be addressed, and the associated metrics or models that can be developed to support BoDs.

This systematic literature review (SLR) is structured as follows. It first provides the research background including the research problem. It then discusses the research method used and results obtained. Finally, it outlines key insights and contributions before concluding.

## Research Background

An analysis of the term "cybersecurity" revealed a range of definitions in dictionaries, academic literature, and industry sources depending on context. A common theme though that runs across the dictionaries defines cybersecurity as "measures taken to protect a computer or computer system (as on the Internet) against unauthorized access or attack" (Merriam, W. 2013), but with no definite consensus. This high variability in definition indicates the absence of a single, unified, inclusive definition that captures all aspects of cybersecurity. While this is not a critical gap, the lack of a precise, consistent definition of the concept and its vague scope leaves room for interpretation by the multitude of role players in this field. It has to be acknowledged in this context that there is a need to develop concise and universally acceptable definition will enable a clearer understanding and application of cybersecurity governance.

While the non-standardised cybersecurity terminology may lead to erroneous focus or interpretation, as well as gaps in understanding, there is also a lack of consistency with regards to supporting information. Factors that may negatively affect the strategic decision-making process for cybersecurity and implicitly bring further challenges to the BoDs may be grouped into three main groups: technical, communication, and human processing. From the technical category, the most notable issue is the lack of metrics to assess cybersecurity investments (Gale, Bongiovanni and Slapnicar, 2022) as well as the lack of consistent cybersecurity governance models inclusive of metrics for BoDs. The communication factors include deciding what risk information to report and how to communicate it as well as the existing perception of risk that influence decision-making. Finally, perhaps the most challenging part relates to the human factors, starting with human behaviour and the social and cultural factors in the organization, and ending with understanding at the BoD level the risks from linked failures and determining their financial and reputational impact; this last category has perhaps the most significant and the least quantifiable impact on deciding the institutional level of investment for a cybersecurity company.

# Research Methodology

## *Research Questions*

As stated from the outset, the aim of this study is to investigate the existing risk measurement instruments, cybersecurity metrics, and associated models for supporting BoDs. Within the defined ecosystem, we focused on the following research questions:

*RQ1: Given the challenges that impact BoDs, what data driven metrics are available to provide strategic insight into the cybersecurity governance program of an organization?*

*RQ2: Which of the current models can be used to formulate cybersecurity data-driven metrics so that BoDs can effectively assess the cybersecurity posture of their organization?*

## *Search Criteria*

To answer the above research questions, the approach proposed in (Okoli and Schabram 2010) for identifying the relevant research papers was used followed by the systematic literature review. The first step of the process was to select several relevant information repositories, from academic research to industry documentation. The primary search was performed using the Google Scholar, Science Direct and Springer databases; the IEEE was excluded as it is mostly focused on the engineering aspects of the technologies but not on the governance. In addition, industry sources as shown in Fig.1 were used. The initial search employed a specific set of keywords and key phrases, with additional inclusions and exclusions constraints to identify the relevant sources. The full process of SLR is outlined in Fig. 1.

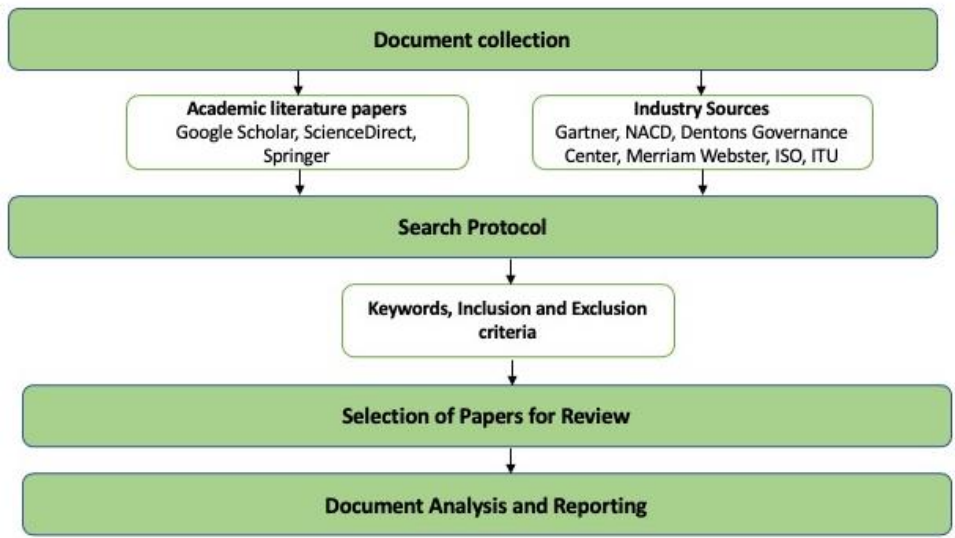

**Figure 1: Summary of research methodology used.**

Based on preliminary research, it became apparent that one of the main challenges that BoDs face as part of their *cybersecurity governance* is identifying *data-driven metrics* and *frameworks* using an accessible *lexicon* to support their decision-making process. As a result, the initial search query focused on a combination of these terms, more specifically the search query below:

`"cybersecurity" AND ("governance" OR "board of directors" OR "data-driven" OR "metrics" OR "frameworks" OR "lexicon")`

A number of Inclusion (IC) and exclusion (EC) criteria for this study were added to improve the relevance and timeliness of the results, as listed below:

- IC1: Filtered search criteria as keywords and key phrases
- IC2: Time period from 1 January 2005 to September 2022
- IC3: Papers presented in English
- IC4: Targeted at BoDs
- IC5: Skimming through the abstract of papers to determine relevance
- EC1: Time period before 2005
- EC2: Any paper targeted at the cybersecurity practitioner or other technical audience

A paper was accepted if it satisfied all of IC1, IC2 and IC3 and one of or both IC4 and IC5, and was excluded if it matched any of the exclusion criteria. We used IC4 and EC2 to set the boundaries in identifying the gap in the literature and, implicitly, in the typical information presented to BoDs. EC1 marks the beginning of formal recognition of cybersecurity by the US government and its security agencies (CISA, 2003).

*Literature Selection Process*

The paper selection process consisted of five filtering stages as shown on Fig.2, each aiming to narrow the scope and the results of the search so that a list of relevant publications are selected for review. The initial search identified 392 papers from academic literature sources along with 66 papers from industry sources. The results from this search were narrowed down by applying further filters on time periods and language, more specifically by selecting only English papers published between 2005 and 2022. Following manual analysis, duplicate papers were removed and the selection was further reduced by skimming through the abstract sections to determine relevance and discarding the irrelevant publications. Finally, an additional set of six papers were added through the snowballing method, which were relevant for the research question, and did offer useful background relating to the topic. This led to a total of 21 papers from academic literature and 8 articles from industry sources, that were carried further to the analysis and reporting stage of systematic literature review.

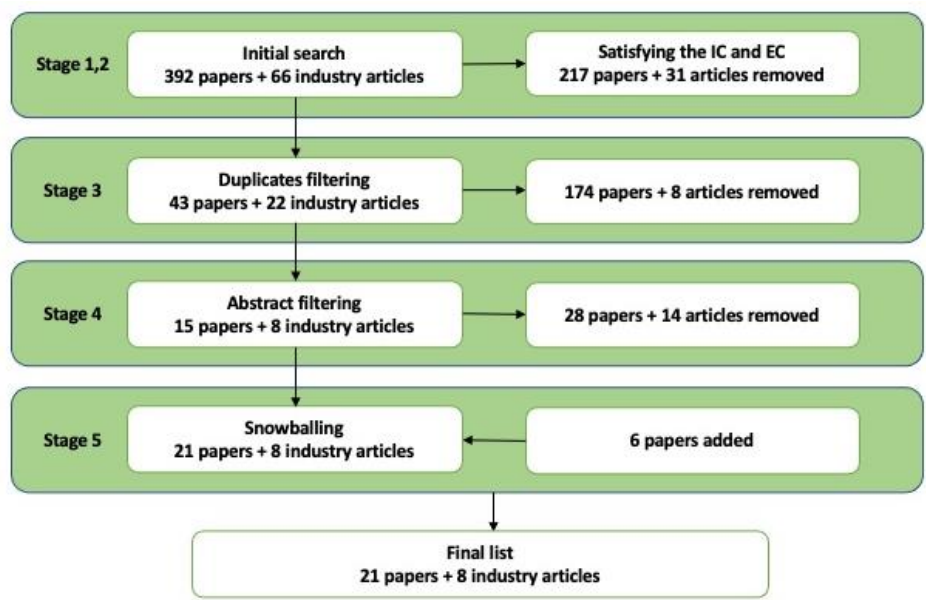

**Figure 2: The paper selection process indicates the papers that were excluded and added, and a final selected set.**

*Data Driven Approaches to Cybersecurity Governance for Board Decision-Making*## Analysis

As acknowledged by the earlier sections, the challenges that BoDs face are paramount and have a direct impact on their ability to take informed decisions. From the spectrum of challenges, this paper specifically focuses on the metrics and models that are available to BoDs and may support them in measuring and governing cybersecurity in their organization.

The shortlisted papers that form the knowledge domain for this study provided a series of useful insights relating to the cybersecurity governance, including both common key elements and distinctive features. These elements and the specific combinations of keywords and phrases were grouped together to derive the following eight thematic categories presented in the Table 1 below.

**Table 1: Conceptual themes identified in the body of reviewed papers and articles.**

| Conceptual Themes | Coverage | List of papers |
|---|---|---|
| Principles | Principles covering oversight, setting the tone, and approach that BoDs may take to improve cybersecurity governance. | (Schinagl and Shahim, 2020), (Donalds, C., and Osei-Bryson, K.-M. 2020), (Maynard, et al., 2018), (Maleh, et al, 2017), (Clinton, L. 2021), (Blum, 2020), (Mandy, C., Olyaei, S., and Proctor, P. 2021), (NACD, 2020) |
| Cybersecurity Lexicon/ Taxonomies | Use of lexicon, taxonomies, and terminologies for BoDs. | (Girn, 2022), (Von Solms, 2016) (AlGhamdi, S., Win, K. T., and Vlahu-Gjorgievska, E. 2020), (Savaş and Karataş, 2022), (Blum, 2020), (William "Bill" Ide III & Leech,2015), (Cybersecurity, n.d.), (Merriam, W. 2013), (NCSC, 2023) |
| Cybersecurity Framework | Cybersecurity governance frameworks/standards in practice | (Girn, 2022), (Von Solms, 2016), (Maynard, et al., 2018), (Maleh, et al., 2017), (Hoong and Rezania, 2022), (Carcary et al. 2016), (ISO, 2018) |
| Accountability | BoDs accountability in managing cyber risks. | Schinagl and Shahim, 2020), (Maynard, Tan, Ahmad and Ruighaver, 2018), (Clinton, L. 2021) (Koh et al., 2005), (Gale, Bongiovanni and Slapnicar, 2022), (Savaş and Karataş, 2022), (Blum, 2020), (William "Bill" Ide III & Leech, 2015), (Proctor, P. 2021), (NACD, 2020) |
| Perception of Cyber Risk | BoDs own perception of their organization's cyber risk | (McFadzean, Ezingeard and Birchall, 2007), (Schinagl and Shahim, 2020), (McLaughlin, M.-D., and Gogan, J. 2018), (Marchewka, 2022) (Soomro, Shah and Ahmed, 2016), (Rocha Flores, Antonsen and Ekstedt, 2014), (Blum, 2020), (Proctor, P. 2021), (NCSC, 2023) |
| Culture | BoDs role in fostering a culture of cybersecurity awareness. | (McFadzean, Ezingeard and Birchall, 2007), (Schinagl and Shahim, 2020), (Donalds, C., and Osei-Bryson, K.-M. 2020), (Maynard et al., 2018), (Kayworth and Whitten, 2022) ,(Koh et al., 2005), (Soomro, Shah and Ahmed, 2016), (Rocha Flores, Antonsen and Ekstedt, 2014), (William "Bill" Ide III & Leech, 2015), (NCSC, 2023) |
| Reporting/ Communication | Risk reporting communication between BoDs and management to support the execution of cybersecurity governance responsibilities. | (Girn, 2022), (Soomro, Shah and Ahmed, 2016), (Savaş and Karataş, 2022), (Carcary et al. 2016), (Nottingham, 2014), (Scholtz, T. 2021) |

The remainder of this section will individually discuss the conceptual themes.

*Principles*

In the context of cybersecurity, the underlying concepts and principles have been identified by prior research as being one of the pillars of effective governance. As a result, there are various literature and industry sources that provide extensive guidance on targeted at BoDs. A typical example of such efforts is the Director's Handbook On Cyber-Risk Oversight, an annual review conducted in recent years for the National Association of Corporate Directors (NACD) by the Internet Security Alliance. The handbook provides a detailed briefing of the cybersecurity landscape and its relationship with and implications on the wider business, as well as a section with critical questions and useful actions for directors to action to "understand their organization's current position, exercise their oversight function and set future goals" (NACD, 2023). According to (Blum, 2020), "boards of many larger companies in regulated industries are formally instituting these kinds of practices" and as a result "seeing an increase in Board accountability and awareness for cybersecurity". In a review conducted by (Clinton, 2020) he further stated that, according to an earlier survey conducted by PricewaterhouseCoopers (PwC), the use of the handbook was associated with "higher cyber security budgets, better cyber risk management, closer alignment of cyber security with overall business goals and helping to create a culture of security within the organizations that adopted it". This direct implication demonstrates that there is adequate guidance on principles to assist BoDs to govern cybersecurity, which is both beneficial for the business sector but requires support and cooperation from all stakeholders.

*Cybersecurity Lexicon and Taxonomies*

An analysis of the existing publications revealed that the cybersecurity governance lexicon and terminology differ vastly depending on the target audience, whether they are aimed at BoDs or technical audiences. While technical taxonomies do exist, appropriate and supportive guidance for BoDs in a language, granularity, and style they can understand and apply is rather limited. The issue stems from the content of the well-known frameworks such as ISO and NIST, which need to be translated from technical terminology to a language that BoDs can understand and relate to. Whilst some taxonomies do exist such as the Factor Analysis of Information Risk (FAIR™) that can be applied to some areas of cyber risk, reviews state that the complexity of the associated frameworks typically relies on the use of automated software and "confuses decision-makers who have no thorough understanding of technology" (RSI Security, 2021). For an optimal outcome, cybersecurity governance requires an integrated data-driven model for BoDs consisting of a clear, consistent lexicon, taxonomy and metrics that is universally accepted.

*Cybersecurity Frameworks*

There is a substantive volume of guidance aiming to assist in governing the cybersecurity posture of organizations in terms of frameworks. Well-known frameworks include the globally recognised International Organization of Standards (ISO) standards such as the 27001 and 38500 series, the National Institute of Standards and Technology (NIST) standards such as NIST Cybersecurity Framework, and the Control Objectives for Information and related Technology (COBIT) (Kuzminykh et. Al., 2021). Prior studies, such as (Freund and Jones, 2014), provided a comparative review of these frameworks, including financial and legal impact but developed primarily from a technical perspective, as summarised in Fig. 3. While these are good practices and are widely supported, they are prescriptive documents, designed for compliance, and aimed at a cybersecurity technical audience rather than the BoD governing the organization. Regarding the ISO27000 standard series, "little information is given about security objectives, potential implementation strategies for these objectives or about the key aspect of accountability arrangements" (Maynard et al. 2018). In an article by Schingal and Shahim, they also allude to the point that the "contextual security governance challenges that an organization faces" are not taken into consideration and that a "paradigm shift is required to move from internally focused protection of organization-wide information towards an embedded and resilient view that considers an organization's collaborative business environment" (Schinagl and Shahim, 2020). To conclude, it is apparent that cybersecurity governance also requires an underlying descriptive model tailored specifically to assist with the needs of the BoD.

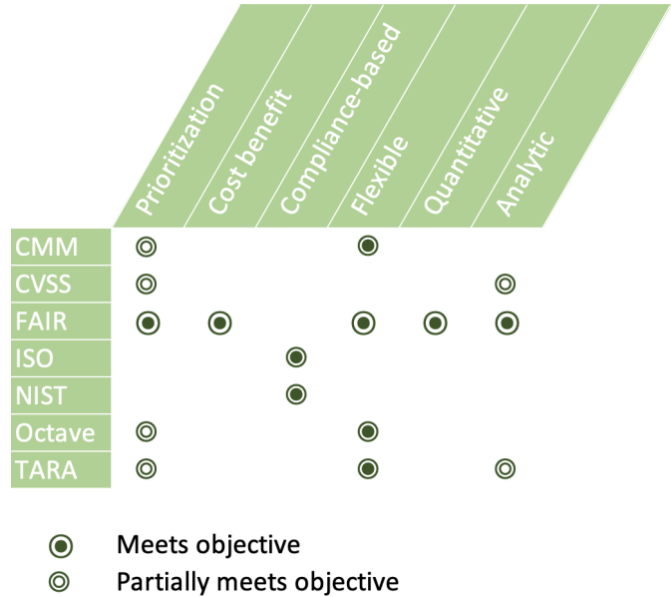

**Figure 3: Comparison analysis of different cybersecurity frameworks.**

## BoDs Accountability

Based on the discussion from the previous sections, it is apparent that, although BoDs bear the corporate governance responsibility for managing cyber risks, they are not equipped with the right knowledge, lexicon, frameworks, and metrics to efficiently exercise their authority and guide them in making informed decisions. In his article, (Girn, 2022) states that "Boards see cybersecurity as top source of risk and as such should ensure they can exercise their accountability with the right skills themselves and ensure reporting into committees engages in the right language". The study also suggests that accountability for cybersecurity should extend beyond the BoD to the leadership teams and to the organizational structures. The same study indicates that "centralized IT governance with a framework of indicators and metrics for reporting, gives a better outcome for reducing cybersecurity risk". Despite the acceptance of cybersecurity governance accountability, (Von Solms, 2016) points out that there is "no real comprehensive (self-assessment driven) maturity model exist which can help a Board to self-assess their level of cyber accountability". Without effective oversight and accountability, an organization's cybersecurity governance systems, policies, and procedures can be rendered meaningless, leaving the enterprise vulnerable to attack (William "Bill" Ide III & Leech, 2015). To summarise this perspective, BoDs accept their corporate governance accountability to manage cybersecurity but do not have a universally accepted maturity model to help them self-assess their knowledge and understanding of the impact of cyber risks to their organization or the accuracy and effectiveness of a cybersecurity strategy implementation.

*Perception of Cyber Risk*

This category focuses on how BoDs perceive cybersecurity and whether their perception influences their decision-making process while approving a cybersecurity strategy and overseeing its implementation. A study by (McFadzean, Ezingeard and Birchall, 2007) determined that the perception of risk by a BoD and its understanding of the role of information systems "influences the planning, adoption and use of organizational security". The study argues that, despite existing research on grounded theories in this area, there is a gap in determining the set of critical attributes that determines the perception of cybersecurity risk by a BoD and how these perceptions influence their decision-making regarding cybersecurity issues. In another paper that tackled the perception of risk and the strategic impact of existing IT on cybersecurity governance at board level, (Marchewka, 2022) points out that "risk perception is a crucial part of risk decision making". Further, the study claims that "a change in perception may reduce the potential losses to organizations and customers depending on the actions taken", highlighting the necessity for improving and evaluating the perception of risk by BoDs (Marchewka, 2022). Ultimately, there is a clear need for research on optimising the communication techniques to improve awareness and perception of cybersecurity risks.

*Culture*

A wide range of articles covered the influence that cybersecurity governance can have on the culture within an organization. (Koh et al., 2005) points out that, in order to institutionalise proper user behaviour towards information security within an organization, a security culture has to be created. The importance of users' roles in effective security management and a good security culture is not using technology but rather the human element, and can be achieved through such activities as leadership, clear communication, and training (NCSC, 2023). There is not enough research on "how to create a good security culture or even what constitutes a good security culture" (Koh et al., 2005). The connectivity and relationship between technology and the human element is critical to the success of the process, fact highlighted by (William Bill Ide III & Leech, 2015), where it is emphasised that cybersecurity "requires a culture of accountability, collaboration and continuous education and training, with all efforts geared toward supporting the strategy and mitigating cyber risks". This shared commitment starts with the BoDs leading by example and the BoDs oversight role in ensuring that cybersecurity programs must extend to the cultural values of the organization; prevailing to do so potentially creates a conflict between the demands of the security program and the values of the firm, leading to employee resistance of security policies. The success of the efforts to improve security culture and their effectiveness must further be qualitatively measured through frequent communications, awareness campaigns, and quantitatively, using the attendance to awareness training and the completeness rates of awareness training courses. These metrics, while they are an indicator for progress, they do not guarantee success as they only verify the user attitude and behaviour towards security rather than the wider cultural elements. To conclude, a cybersecurity cultural toolkit should be present, specifically targeted at BoDs to drive cyber risk awareness and, to complete the landscape, industry standards for developing metrics to include a broader set of cultural elements are also necessary.

*Reporting / Communication*

Effective cyber risk communication between the BoDs and management is critical and expectations for effective cyber risk management have increased over the last decade. In his article, (Nottingham, 2014) states that "regulators and investors all demand that BoDs have effective cyber risk management and communication systems in place, while credit rating agencies have also expanded their review of risk practices in determining overall creditworthiness". Further, the report of Oliver Wyman's Global Risk Center argues that BoDs "remain dissatisfied with the current state of risk communication." The study points out that, according to a survey conducted by the NACD, half of all respondents believe that "improvements in risk communication are needed". Without effective risk communication, BoDs fail to provide the necessary oversight and guidance, therefore "unfortunately, many BoDs are frustrated by the mismatch between the risk information that is provided to the board and the information that is requested" (Nottingham, 2014). Clarity on the nature of reporting required to enable BoD accountability is limited. Gartner (Scholtz, 2021) also states that "security presentations and reports do not resonate with senior leaders and the Board and are rarely aligned to business drivers; in fact, security is seen as a necessary evil rather than a business investment". BoDs therefore require more effective cyber risk management and communication systems in place to support them to govern cybersecurity.

## Discussion

Considering the themes discussed above, the in-depth analysis of the papers in this review demonstrated that there is a lack of tangible and meaningful metrics and models to govern the cybersecurity information available to the BoDs. While some guidance and support exist, they are mostly aimed at a technical audience and not targeting the BoDs. This finding overlaps with our previous study in the area of IOT security for business that revealed a poor uptake and management of the environment due to lack of standards, appropriate metrics in cybersecurity, or support from top management due to not understanding the security aspects (Kuzminykh, Ghita & Such, 2022). The Common Ground theory that refers to "mutual knowledge, mutual beliefs, and face-to-face communication" (Clark and Brennan, 1991) can be used to translate technical language from existing cybersecurity lexicon to develop business-like lexicon for BoDs in a language they can understand and in educating senior company leaders (Dube and Mohanty, 2022).

This review also identifies the need for developing a holistic framework for cybersecurity governance that not only aligns the objectives of an organization and its security strategy but also embeds organizational and security culture into the framework. A number of frameworks to define metrics have been proposed by prior studies. One such framework is the CARE framework (Mandy et al. 2021), using a model proposed by Gartner. This framework categorises security metrics into consistency and reasonableness of security controls, adequacy of the security program, and effectiveness of security resources. This framework could be useful to formulate more meaningful metrics for BoDs to assist specifically with making decision to govern cybersecurity. These metrics are also paramount to effective reporting and communication between cybersecurity technical audiences and BoDs.

The Cyber Security Maturity Model for Boards of Directors (CSMMBoD), developed by (Von Solms, 2016) can help to determine the cybersecurity maturity level of BoDs to help prepare them better for their accountabilities. The proposed model is designed as a self-assessment exercise and assesses the BoDs knowledge and their understanding of the impact of cyber risks to their organization, the execution of the cybersecurity strategy, policies and technologies, the organizational structures, and the budget of the organization. This model could be further developed to include elements of social and human behaviour and organizational cultural responsibilities for cybersecurity governance.

In order to provide an a priori understanding of the preparedness from a human factor perspective, theories such as Theory of Planned Behaviour (TPB) use a predictive model that indicates that subjective norms and attitudes influence behavioural intention and hence actual behaviour of humans. Until now, insufficient research has been done on subjective norms regarding cybersecurity and TPB can be useful to understand the roles of various behavioural factors and identifying the ones that will have the highest predictive value to integrate it in a decision-making plan, one required for BoDs. Social Cognition Theory (SCT) indicates that there is a reciprocal cause and effect between a person's behaviour and both the social world and individual characteristics. SCT can be applied to cybersecurity to investigate decision support and behaviour. Further research into these models could be useful when integrating human and social factors and organizational cultural factors into the cybersecurity framework.

As discussed, cyber risk perception is a crucial contributor that influences decision-making. Based on a study observing the perception of risk and the strategic impact of existing IT on information security strategy at board level, (McFadzean, Ezingeard and Birchall, 2007) proposes the Perception Grid Model to help BoDs to understand the social issues, perception of risk, human behaviour patterns within cybersecurity in a more holistic way. Another tool that can be used to "break the ice" in the understanding of importance of cybersecurity metrics, activities, and indicators of success, is the Cyber Security Toolkit for Boards (NCSC, 2023). The toolkit consists of three parts: background information, nine core modules, presented in Fig. 4, and a set of relevant cyber security regulations that BoDs need to be aware of. Further research into the model could be conducted to determine if it closes the gap between how directors perceive their organizational security and what factors influence their decisions on governing cybersecurity.

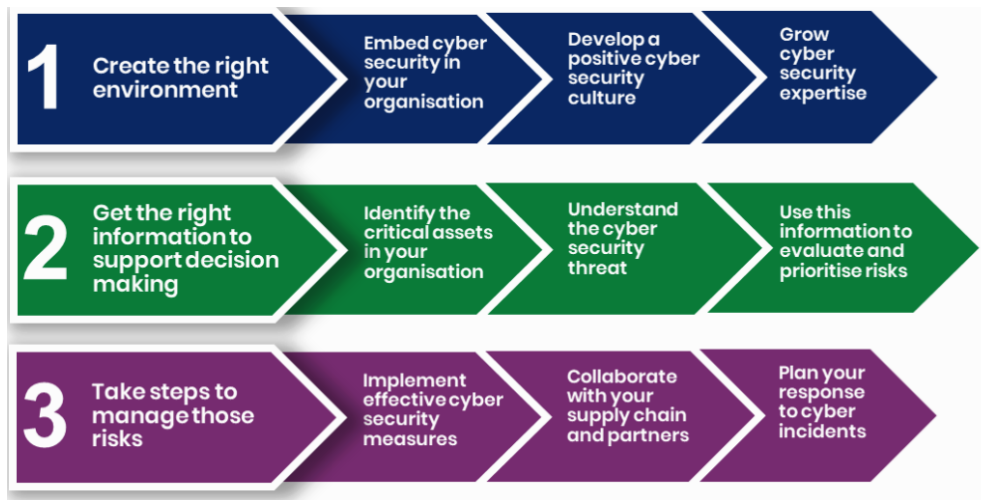

**Figure 4: Nine key cyber security topics grouped into three categories that make up the core of the Cyber Security Toolkit for Boards (NCSC, 2023).**

The findings and recommendations to the research questions are highlighted in Table 2.

**Table 2: Summary for findings and recommendations**

| Research Question | Findings |
| --- | --- |
| RQ1: Given the challenges that impact BoDs, what data driven metrics are available to provide strategic insight into the cybersecurity governance program of an organization? | 1) There is a lack of metrics and models that are tangible and meaningful to govern cybersecurity available to the BoDs. <br><br> 2) While some guidance and support are available, this is mostly to a technical audience and not targeted at the BoDs in a language they understand. |
| RQ2: Which of the current models can be used to formulate cybersecurity data-driven metrics so that BoDs can effectively assess the | 1) A substantial number of guidelines and textbooks exists and may be used to explain the general principles of effective governance <br><br> 2) The Grounding communication theory. (Clark and Brennan, 1991) is a good model to use for the Cybersecurity lexicon and taxonomies |

| cybersecurity posture of their organization? | 3) Further development of the CARE model (Mandy et al. 2021) is required for formulating appropriate metrics within Cybersecurity frameworks |
| --- | --- |
| | 4) Self-assessment capability maturity models could help evaluating BoDs and providing them with the necessary accountability |
| | 5) The Theory of Planned Behaviour (TPB) and Social Cognition Theory (SCT) can be utilised to define organisation-cultural metrics |
| | 6) The Perception Grid Model (McFadzean, Ezingeard and Birchall, 2007) or the The Cyber security Board toolkit may be employed to explain the metrics for Cybersecurity risk |
| | 7) Reporting and communication metrics could be further simplified through additional development of the CARE model (Mandy et al. 2021) |

## Conclusions

Understanding how certain aspects of cybersecurity governance influence the quality of strategic decision-making in cybersecurity is an essential step to ensuring that cyber risks are managed and that security investments are used effectively. BoDs are ultimately accountable for making these strategic decisions; however, they are not equipped with meaningful data-driven metrics and useful models that is consistent, standardised, universally approved in industry and in a language they understand.

This study reviewed 21 academic papers and 8 industry papers to find suitable metrics and models for BoDs to assist them in cybersecurity decision making. The findings revealed that there is limited academic research and industry sources aimed at the BoD audience. A number of recommendations into theories and models to formulate these metrics have been made to improve the current ecosystem and future research in the area should focus on the impact on existing and proposed method on the effectiveness and accountability of BoDs in the context of cybersecurity governance.

## References

AlGhamdi, S., Win, K. T., Vlahu-Gjorgievska, E. (2020). Information Security Governance Challenges and Critical Success Factors: Systematic Review. Computers & Security, 99, 102030. https://doi.org/10.1016/j.cose.2020.102030

Blum, D. (2020). Manage Risk in the Language of Business. Rational Cybersecurity for Business, 123–156. https://doi.org/10.1007/978-1-4842-5952-8_5

Carcary, M., Renaud, K., McLaughlin, S., & O'Brien, C. (2016). A Framework for Information Security Governance and Management. IT Professional, 18(2), 22–30. https://doi.org/10.1109/mitp.2016.27

CISA, Cybersecurity and Infrastructure Security Agency (2003, Dec 17). *Homeland Security Presidential Directive 7*. https://www.cisa.gov/news-events/directives/homeland-security-presidential-directive-7

Clark, H. H., Brennan S.E. (1991). Grounding in communication. In Lauren Resnick, Levine B., M. John, Stephanie Teasley & D. (eds.), Perspectives on Socially Shared Cognition. American Psychological Association. pp. 127--149.

Cybersecurity. (n.d.). Www.itu.int. https://www.itu.int/en/ITU-T/studygroups/com17/Pages/cybersecurity.aspx#:~:text=Cybersecurity%20is%20the%20collection%20of

Clinton, L. (2021). International principles for boards of directors and cyber security. Cyber Security: A Peer-Reviewed Journal, 4, 243–250. http://isalliance.org/wp-content/uploads/2021/04/CSJ_4_3_CSJ0005_Clinton.pdf

Disterer, G. (2013). ISO/IEC 27000, 27001 and 27002 for Information Security Management. Journal of Information Security, 04(02), 92–100. https://doi.org/10.4236/jis.2013.42011

Donalds, C., & Osei-Bryson, K.-M. (2020). Cybersecurity compliance behavior: Exploring the influences of individual decision style and other antecedents. International Journal of Information Management, 51, 102056. https://doi.org/10.1016/j.ijinfomgt.2019.102056

Dube, D.P. and Mohanty, R.P. (2022). Application of grounded theory in construction of factors of internal efficiency and external effectiveness of cyber security and developing impact models. Organizational Cybersecurity Journal: Practice, Process and People. https://doi.org/10.1108/OCJ-04-2022-0009

Freund, J. and Jones, J. (2014). Measuring and Managing Information Risk: A FAIR Approach. Butterworth-Heinemann, ISBN: 9780127999326

Gale, M., Bongiovanni, I., & Slapnicar, S. (2022). Governing cybersecurity from the boardroom: Challenges, drivers, and ways ahead. Computers & Security, 121, 102840. https://doi.org/10.1016/j.cose.2022.102840

Gartner, An Outcome-Driven Approach to Cybersecurity Improves Executive Decision Making. (n.d.).. Retrieved March 2, 2023, from https://www.gartner.com/en/documents/3980893

Gartner, Capture and Communicate The Value of Cybersecurity. (n.d.). Retrieved March 2, 2023, from https://www.gartner.com/en/doc/value-of-cybersecurity

Gartner, Metrics to Prove You CARE About Cybersecurity. (n.d.). Retrieved March 2, 2023, from https://www.gartner.com/en/documents/4003605

Girn, S. (2022). A Data Driven Approach to Board Cybersecurity Governance. Proceedings of Pacific Asia Conference On Information Systems (PASIC). http://hdl.handle.net/10453/159073

Hoong, Y., & Rezania, D. (2022). Cybersecurity Governance in Information Technology: A Review of What Has Been Done, and What Is Next. Computer Networks, Big Data and IoT, 285–294. https://doi.org/10.1007/978-981-19-0898-9_22

Kayworth, T., & Whitten, D. (2010). Effective Information Security Requires a Balance of Social and Technology Factors. Ssrn.com. https://papers.ssrn.com/sol3/papers.cfm?abstract_id=2058035

Koh, K., Ruighaver, A., Maynard, S., Ahmad, A. (2005). Security Governance: Its Impact on Security Culture.. Proceedings of 3rd Australian Information Security Management Conference. 47-58

Kuzminykh I, Ghita B, Sokolov V, Bakhshi T. (2021). Information Security Risk Assessment. Encyclopedia 2021;1:602–17. https://doi.org/10.3390/encyclopedia1030050

Kuzminykh, I., Ghita, B., Such, J.M. (2022). The Challenges with Internet of Things Security for Business. In: Koucheryavy, Y., Balandin, S., Andreev, S. (eds) Internet of Things, Smart Spaces, and Next Generation Networks and Systems. NEW2AN ruSMART 2021 2021. Lecture Notes in Computer Science(), vol 13158. Springer, Cham. https://doi.org/10.1007/978-3-030-97777-1_5

Maleh, Y., Ezzati, A., Sahid, A., Belaissaoui, M. (2017). CAFISGO: a Capability Assessment Framework for Information Security Governance in Organizations. Journal of Information Assurance & Security. 2017, Vol. 12 Issue 6, p209-217. 9p.

Mandy, C., Olyaei, S., and Proctor, P. (2021). "Metrics to Prove You Care About Cybersecurity," Gartner.

Marchewka, Edward. (2022). Reducing Cybersecurity Risk Information Asymmetry Phenomenon: A Prescriptive Approach to Improving Cybersecurity Risk Perception. 10.13140/RG.2.2.26205.49125

Maynard, S. B., Tan, T., Ahmad, A., & Ruighaver, T. (2018). Towards a Framework for Strategic Security Context in Information Security Governance. Pacific Asia Journal of the Association for Information Systems, 65–88. https://doi.org/10.17705/1pais.10403

McFadzean, E., Ezingeard, J., & Birchall, D. (2007). Perception of risk and the strategic impact of existing IT on information security strategy at board level. Online Information Review, 31(5), 622–660. https://doi.org/10.1108/14684520710832333

McLaughlin, Mark-David and Gogan, Janis (2018) "Challenges and Best Practices in Information Security Management," MIS Quarterly Executive: Vol. 17: Iss. 3, Article 6. Available at: https://aisel.aisnet.org/misqe/vol17/iss3/6 Secure File Repository. (n.d.). Www.nacdonline.org. Retrieved March 2, 2023

Merriam-Webster, Definition of CYBERSECURITY. (2019). Merriam-Webster.com. https://www.merriam-webster.com/dictionary/cybersecurity

NACD Director's Handbook on Cyber-Risk Oversight. (n.d.). Www.nacdonline.org. https://www.nacdonline.org/insights/publications.cfm?ItemNumber= 74777

NCSC, National Cyber Security Center. (2023, March 30). *Cyber Security Toolkit for Boards*. https://www.ncsc.gov.uk/collection/board-toolkit/introduction-to-cyber-security-for-board-members

Okoli, C. and Schabram, K. (2010). A Guide to Conducting a Systematic Literature Review of Information Systems Research. http://dx.doi.org/10.2139/ssrn.1954824

Pescatore, J., Spitzner, L. (2017, Oct 18). Security Awareness for Board of Directors. SANS webcast. https://www.sans.org/webcasts/security-awareness-board-directors-105920/

Proctor, P. (2021). An Outcome-Driven Approach to Cybersecurity Improves Executive Decision Making, Gartner.

Rocha Flores, W., Antonsen, E., & Ekstedt, M. (2014). Information security knowledge sharing in organizations: Investigating the effect of behavioral information security governance and national culture. Computers & Security, 43, 90–110. https://doi.org/10.1016/j.cose.2014.03.004

RSI Security (2021). Pros and Cons of Factor Analysis of Information Risk. https://blog.rsisecurity.com/pros-and-cons-of-factor-analysis-of-information-risk

Savaş, S., & Karataş, S. (2022). Cyber governance studies in ensuring cybersecurity: an overview of cybersecurity governance. International Cybersecurity Law Review. https://doi.org/10.1365/s43439-021-00045-4

Schinagl, S., & Shahim, A. (2020). What do we know about information security governance? Information & Computer Security, 28(2), 261–292. https://doi.org/10.1108/ics-02-2019-0033

Scholtz, T. (2021). How to Communicate the Value of Information Security in Business Terms. Gartner.

Soomro, Z. A., Shah, M. H., & Ahmed, J. (2016). Information security management needs more holistic approach: A literature review. International Journal of Information Management, 36(2), 215–225. https://doi.org/10.1016/j.ijinfomgt.2015.11.009

von Solms B. (2016). Towards a cyber governance maturity model for boards of directors. The Business and Management Review, Vol. 7, no. 4. https://cberuk.com/cdn/conference_proceedings/conference_36136.pdf

William Bill Ide III, R. and Leech, A. (2015) A cybersecurity guide for directors. Denton's governance: Course hero: https://www.coursehero.com/file/72499064/cybersecurity-090115pdf/

Nottingham, L. (2014). Risk communication: Aligning the board and c-suite. Report Oliver Wyman's Global Risk Center. https://www.oliverwyman.com/our-expertise/insights/2014/feb/risk-communication-2014.html